\documentclass[aps,prb]{revtex4}

\usepackage{graphicx}
\usepackage{dcolumn}
\usepackage{bm}

\begin{document}

\title{Instability and decomposition on the surface of strained alloy films}
\author{Zhi-Feng Huang}
 \email{zfh@physics.utoronto.ca}
\author{Rashmi C. Desai}
 \email{desai@physics.utoronto.ca}
 \homepage{http://www.physics.utoronto.ca/people/faculty/desai.html}
\affiliation{%
Department of Physics, University of Toronto, \\
Toronto, Ontario, Canada M5S 1A7
}%
\begin{abstract}
A continuum dynamical model is developed to determine the 
morphological and compositional instabilities on the free
surface of heteroepitaxial alloy films in the absence of 
growth. We use linear stability analysis to study the early 
nonequilibrium processes of surface evolution, and calculate 
the stability conditions and diagrams for different cases of
material parameters. There are two key considerations in our
treatment: the coupling between top free surface of the film
and the bulk phase underneath, and the dependence of both 
Young's and shear elastic moduli on local composition. The 
combination and interplay of different elastic effects caused 
by lattice misfit between film and substrate (misfit strain), 
composition dependence of film lattice parameter (compositional 
strain) and of film elastic constants lead to complicated and 
rich stability results, in particular the joint stability or 
instability for morphological and compositional profiles, the 
asymmetry between tensile and compressive layers, as well as 
the possible stabilization and suppression of surface 
decomposition even below the effective critical temperature.
We also compare our results with the observations of some
postdeposition annealing experiments.
\end{abstract}



\maketitle

\section{Introduction}

Instability is one of the most important phenomena in strained
heteroepitaxial growth, where a thin film of different material
is epitaxially deposited on a substrate and the mismatch of
lattice constants between film and substrate leads to strain
in the grown film. Besides the formation of misfit dislocations 
and other defects, the strain can be relieved through the process 
of morphological instability, during which the growth mode of 
the film is changed from layer-by-layer to three-dimensional
island but the film still remains coherent with its substrate.
This dislocation-free morphological instability occurs in a wide 
range of heteroepitaxial systems, and has been well investigated 
both theoretically \cite{asaro,grinfeld,srolovitz,spencer91} and 
experimentally \cite{legoues,eaglesham} for single-component 
materials .

In recent years, more attention has been paid to the 
multi-component strained layers, in particular the binary 
(e.g., Si$_{1-x}$Ge$_x$ \cite{xie,walther,perovic,lagally,
tromp}) and pseudo-binary (e.g., In$_{1-x}$Ga$_x$As 
\cite{snyder,gendry,guyer00,okada,peiro,gonzalez}) alloys. 
Since the materials are composed of different kinds of atoms 
that may not be fully miscible, the compositional 
inhomogeneities are expected to develop at certain growth 
conditions. This alloy segregation instability couples with 
the morphological instability, resulting in the simultaneous 
modulations of the surface profile and alloy composition 
during the epitaxial growth.\cite{walther,okada,peiro,gonzalez}
Another important phenomenon of the heteroepitaxial alloy 
films is the asymmetry of the stability for compressive
and tensile layers. For Si$_{1-x}$Ge$_x$, films grown under
compression were observed to be less stable than those under
tension,\cite{xie} while for In$_{1-x}$Ga$_x$As the situation
is more complicated: Whether compressive or tensile films
are more stable seems to depend on the material's growth rate.
\cite{gendry,guyer00,okada} Compared with single-component
films, there is an extra effect on alloy strain due to 
composition dependence of the film lattice constant.
\cite{cahn,larche} The combination of this compositional 
strain and the film-substrate misfit strain highly 
influences the stability behaviors of alloy films.
\cite{guyer96,spencer00}

Some theoretical investigations have been carried out in
order to understand the morphological instability of 
strained multi-component films, and most of them focused 
on the properties and behaviors of growing layers with
deposition flux.\cite{guyer96,spencer00,spencer01,
tersoff,leonard98,leonard99,huang} However, the studies
of the static film without deposition \cite{glas,leonard98} 
are much fewer, although the stability of epitaxial
films in the absence of growth is also important
from both the theoretical and experimental points of view.
Using the treatment of thermodynamics, Glas \cite{glas} has
demonstrated that any strained alloy with a static 
free surface is unstable due to the interplay of morphological 
and compositional instabilities. Furthermore, as also
pointed out by Glas, this result is valid provided all the
evolution mechanisms are possible, and when the physically 
relevant mechanism is considered, the unstable state derived
thermodynamically may not be kinetically accessible. This has
been verified in the nonequilibrium and dynamical analysis of
L\'{e}onard and Desai,\cite{leonard98} where the stability
properties of both the non-growing (static) and growing 
strained alloy films were determined. Stabilization in 
non-growing film was found to be possible for some parameter 
values, showing the significance of kinetic evolution process.

In this paper we apply a nonequilibrium, continuum model in
order to study the morphological instability and surface 
decomposition of binary or pseudo-binary strained non-growing 
alloy films, i.e., films without deposition. The system is 
assumed to be elastically isotropic and follow the conserved 
dynamics of phase separation and surface diffusion. Compared 
with the model of L\'{e}onard and Desai\cite{leonard98} and 
other previous work, here we have two crucial considerations 
which lead to significantly different stability results. First, 
due to the fact that the surface phase and bulk phase in a 
strained film are intimately coupled with each other, we take 
into account the total free energy of the system, not just the 
surface state and energy as used before.\cite{leonard98} Second, 
the film elastic moduli (both the Young's modulus $E$ and shear 
modulus $\mu$) are considered to be composition-dependent, 
which deeply affects the stability properties. Presenting the 
linear analysis on early evolution of surface profile and 
composition, we derive the stability conditions for non-growing 
film as well as stability diagrams for various material parameters 
appropriate to realistic systems.

In Sec. \ref{secII}, we describe the continuum model with 
conserved dynamics and present the mechanical equilibrium 
equation for the elasticity of this film system. The linear 
stability analysis is carried out in Sec. \ref{secIII}, 
resulting in the characteristic equation for perturbation 
growth rate. The results related to the stability conditions 
and diagrams are shown in Sec. \ref{secIV}, and the discussion 
and conclusions are presented in Sec. \ref{secV}.

\section{Model and elasticity formulation}
\label{secII}

We consider a strained alloy system composed of a semi-infinite
substrate occupying the region $z<0$ and a A$_{1-X}$B$_X$ binary
or pseudo-binary alloy film in the region $0<z<h(x,y)$, where
$h(x,y)$ represents the surface height variable. The 
film-substrate interface located at $z=0$ is assumed to be planar
and remain coherent without generating any misfit dislocation. 
In this heteroepitaxial system, the misfit strain caused by the
difference between the lattice constant of the film $a_f$ and 
that of the substrate $a_s$ is characterized by 
$\epsilon=(a_f-a_s)/a_s$. Thus, $\epsilon >0$ or $<0$ implies
a strained film under compression or tension, respectively.
To describe the composition profile of the film, we use a 
continuous variable $\phi({\bf r},t)$ which is defined only 
for $z\geq 0$ and is proportional 
to the local difference in the concentrations of two 
constituents. Corresponding to an alloy composition $X$, its 
average value $\bar{\phi}$ is equal to $2X-1$. Here we focus 
on the symmetric mixture, i.e., $X=1/2$ alloy, for which the 
spinodal decomposition theory can be well applied, and then 
we have $\bar{\phi}=0$ in what follows. Due to the atomic size 
difference between $A$ and $B$ species, the film lattice 
parameter $a_f$ is composition-dependent and the solute 
expansion coefficient $\eta=\partial \ln a_f / \partial \phi$ 
\cite{cahn,larche} is defined to measure the compositional 
strain.

Assume that the thin film has been grown under ultra-high 
vacuum condition, e.g., in a molecular-beam epitaxy (MBE) 
system, and then the evaporation and re-condensation on film 
surface are negligible. Furthermore, we also neglect the 
interdiffusion between film and substrate as well as the 
diffusion and compositional relaxation in the bulk film, 
since the bulk atomic mobility is much smaller than the 
mobility at the surface in typical epitaxial growth. Thus, 
the dynamics of morphological and compositional evolution is 
dominated by the surface diffusion and surface decomposition 
processes, and should be conserved. 

For the evolution of surface profile, the surface diffusion 
mechanism leads to
\begin{equation}
\frac{\partial h}{\partial t} = \Gamma_h \sqrt{g} \nabla_s^2 
\frac{\delta {\cal F}}{\delta h},
\label{eq-h}
\end{equation}
while to measure the time-dependence of concentration field
at the surface $\phi(x,y,h(x,y),t)$$=\phi_s(x,y,t)$, we apply
the conserved dynamics:
\begin{equation}
\frac{\partial \phi}{\partial t} = \Gamma_{\phi} \nabla^2 
\frac{\delta {\cal F}}{\delta \phi}.
\label{eq-phi}
\end{equation}
Note that in Eq. (\ref{eq-phi}) there are two ways to study 
the surface composition fluctuations. The first one
\cite{leonard98} is evaluating the free energy ${\cal F}$ at 
the surface and then calculating the functional differentiation 
with respect to surface composition field $\phi_s$, that is, 
only considering the surface state. Here we take into account 
the intimate coupling between surface state and bulk state and 
use the other way: \cite{huang} Apply the total free energy 
${\cal F}$ of the whole system to calculate the composition 
dynamics and then evaluate it at the surface, as has been done 
in the previous study of surface critical phenomena for spin 
fluctuations.\cite{kumar}

In Eqs. (\ref{eq-h}) and (\ref{eq-phi}), $\nabla_s^2$ is the
surface Laplacian, $g=1+|\nabla h|^2$ represents the 
determinant of the surface metric, and the kinetic coefficients
are denoted as \cite{mullins} $\Gamma_h=D_s N_s/k_B T N_v^2$ 
and $\Gamma_{\phi}=\Gamma_h\delta^{-1}$ with $\delta$ the 
effective diffusion thickness of surface layer. Here $D_s$ 
is the surface diffusivity, $k_B$ is the Boltzmann constant, 
$N_s$ and $N_v$ are the number densities of atoms per unit 
surface area and per unit volume, respectively, and $T$ is the 
temperature. The total free energy functional $\cal{F}$ consists 
of three contributions:
\begin{equation}
\cal{F}=\cal{F}_{\rm s} + \cal{F}_{\rm GL} + 
\cal{F}_{\rm el}.
\label{eq-F}
\end{equation}
The first contribution $\cal{F}_{\rm s}$ is the surface energy, 
which plays a stabilizing role and can be represented by a 
drumhead model without pinning term:
\begin{equation}
{\cal F}_{\rm s}[h]=\gamma \int d^2r \sqrt{g}.
\label{eq-F_s}
\end{equation}
Here $\gamma$ is the surface tension, and for simplicity we
assume it to be isotropic and composition independent. The
second term in r.h.s. of Eq. (\ref{eq-F}) determines the 
phase behaviors of binary compounds and is the 
Ginzburg-Landau functional
\begin{equation}
{\cal F}_{\rm GL}[\phi,h]=\int_{-\infty}^{h} d^3r
\left [-{\frac{r'}{2}}\phi^2 + {\frac{u}{4}}\phi^4
+ {\frac{c}{2}}|\nabla\phi|^2 \right ],
\label{eq-F_GL}
\end{equation}
where the parameters \cite{cahn} $r'=k_B(T_c-T)N_v$, $u$ is 
a temperature independent positive constant, and $c=k_B
T_c N_v a_0^2/2$, with $T_c$ the critical temperature of the 
binary alloy and $a_0$ the effective interaction distance. 
For the bulk alloy without elastic strain, when $T>T_c$ the 
equilibrium state is homogeneous with $\phi=0$, while for 
$T<T_c$ we have the coexistence of two phases $\phi=\pm 
\sqrt{r'/u}$. The last term $|\nabla\phi|^2$ in Eq. 
(\ref{eq-F_GL}) represents the gradient energy that penalizes 
the sharp compositional changes, and is important for stability
analysis: The lack of it leads to a nonphysical divergence 
for short wavelength mode.\cite{guyer96,spencer01}

The last contribution in Eq. (\ref{eq-F}) is the elastic 
free energy functional $\cal{F}_{\rm el}$, and is crucial 
for this stress-driven system. From linear elasticity theory,
it can be expressed as
\begin{equation}
{\cal F}_{\rm el}[\phi,{\bf u},h]=\frac{1}{2}\int_{-\infty}^{h} 
d^3r S_{ijkl}\sigma_{ij}\sigma_{kl},
\label{eq-F_el}
\end{equation}
where $\sigma_{ij}$ is the stress tensor and $S_{ijkl}$ is 
the elastic compliance tensor with the form
$S_{ijkl}=\delta_{ik}\delta_{jl}(1+\nu)/E
-\delta_{ij}\delta_{kl}\nu/E$ for isotropic systems
(subscripts $i$, $j$, $k$, or $l=x$, $y$, $z$). 
Generally, the elastic constants (Young's modulus $E$, 
shear modulus $\mu$, and Poisson ratio $\nu$) are dependent
of the local composition, and here we consider this 
dependence to first order, that is,
\begin{eqnarray}
E&=&E_0(1 + E_1^* \phi), \nonumber\\
\mu&=&\mu_0(1 + \mu_1^* \phi),
\label{eq-Eu}
\end{eqnarray}
and $\nu=E/2\mu-1$. In this paper we take the system as 
being elastically isotropic, and neglect the difference 
in the average elastic constants ($E_0$, $\mu_0$, and
$\nu_0=E_0/2\mu_0-1$) between film and substrate. This
is appropriate for systems with substrate and film having
similar elastic constants.

To determine the elastic energy (\ref{eq-F_el}), we need
to get the solution for the displacement vector ${\bf u}$
which satisfies mechanical equilibrium
\begin{equation}
\partial_j \sigma_{ij}=0
\label{eq-equi}
\end{equation}
in the whole film/substrate system. According to Hooke's 
law for isotropic system, the linear stress tensor is
expressed as
\begin{equation}
\sigma_{ij} = 2\mu \left [\frac{\nu}{1-2\nu}u_{ll}\delta_{ij}
+u_{ij} - \frac{1+\nu}{1-2\nu}(\epsilon +\eta\phi)\delta_{ij} 
\right ],
\label{eq-constit}
\end{equation}
with the presence of misfit strain $\epsilon$ and composition
strain $\eta\phi$, where the linear strain tensor $u_{ij}$ is
given by
\begin{equation}
u_{ij}=(\partial_i u_j + \partial_j u_i)/2.
\label{eq-u_ij}
\end{equation}
The boundary conditions are needed to solve the above mechanical
equilibrium equation. At the free surface of the film, i.e.,
at $z=h(x,y)$, we have
\begin{equation}
\sigma_{ij}^f n_j =0
\label{eq-bound1}
\end{equation}
due to the negligible pressure on the film surface. Here $n_j$
is the unit vector normal to the surface. Since the film-substrate
interface at $z=0$ remains coherent, we get the continuous 
conditions for both stress and displacement tensors:
\begin{equation}
\sigma_{ij}^f =\sigma_{ij}^s
\qquad {\rm and} \qquad u_i^f = u_i^s,
\label{eq-bound2}
\end{equation}
where superscripts $f$ and $s$ refer to the film and substrate,
respectively. Finally, the strains far from the interface, that is,
for $z\rightarrow-\infty$, are expected to decay to zero:
\begin{equation}
u_i^s\rightarrow 0 \qquad {\rm and} \qquad u_{ij}^s\rightarrow 0.
\label{eq-bound3}
\end{equation}

\section{Linear stability analysis}
\label{secIII}

In order to determine the stability properties of this non-growing
strained alloy system, we apply the linear analysis on evolution 
equations (\ref{eq-h}) and (\ref{eq-phi}) with the use of formulae
(\ref{eq-F})--(\ref{eq-bound3}). For a general variable $\xi$, 
which could be displacement $u_i$, composition field $\phi$, or
height variable $h$, its Fourier expansion yields
\begin{equation}
\xi =\bar{\xi} + \sum\limits_{\bf q}\hat{\xi}({\bf q},z,t)
e^{i(q_x x+q_y y)},
\label{eq-expan}
\end{equation}
with small perturbations $\hat{\xi}$ around the basic state 
$\bar{\xi}$ which corresponds to a planar film with fixed 
thickness $h_0=\bar{h}$ and uniform composition $\bar{\phi}=0$.
Note that in Eq. (\ref{eq-expan}) when $\xi$ denotes the height
variable $h$, $\hat{\xi}$ is in fact $\hat{h}({\bf q},t)$.
The basic-state solution \cite{guyer96,leonard98} for the film
leads to $\bar{u}_x^f=\bar{u}_y^f=0$, $\bar{u}_z^f=\bar{u}z$ 
with $\bar{u}=\bar{u}_{zz}^f=\epsilon (1+\nu_0)/(1-\nu_0)$,
$\bar{\sigma}_{xx}^f=\bar{\sigma}_{yy}^f=\bar{\sigma}=
-2\mu_0\bar{u}$, and other stress and strain tensors 
($\bar{u}_{ij}^f$, $\bar{\sigma}_{ij}^f$) to be zero. For the
substrate, the basic state is unstrained and then $\bar{u}_i^s
=0$ and $\bar{u}_{ij}^s=\bar{\sigma}_{ij}^s=0$ ($i$, $j=x$, $y$,
$z$).

The mechanical equilibrium equation (\ref{eq-equi}) with boundary
conditions (\ref{eq-bound1})--(\ref{eq-bound3}) can be solved
to first order $O(\hat{h},\hat{\phi})$ using the above expansion 
(\ref{eq-expan}). Here we use the detailed solutions given by
L\'{e}onard and Desai \cite{leonard98}, where a crucial step is
to introduce a new variable $W$ with
\begin{equation}
\nabla^2 W=\phi,
\label{eq-W}
\end{equation}
or equivalently, $(\partial_z^2-q^2)\hat{W}=\hat{\phi}$ with
$q^2=q_x^2 +q_y^2$. After substituting the solutions in the free
energy functional (\ref{eq-F})--(\ref{eq-F_el}) and then in
the dynamical equations (\ref{eq-h}) and (\ref{eq-phi}), we can
obtain the linearized evolution equations for morphological and
compositional perturbations $\hat{h}$ and $\hat{\phi}$ to 
determine the stability of the system.

What we are interested in is the behavior of perturbations for 
the stressed film without deposition. The fluctuations of alloy
composition mainly occur at the surface $z=h(x,y)$ due to the
surface relaxation process, and should attenuate along the
vertical direction $z$ as the surface/bulk coupling weakens
with the increasing distance from the surface and the bulk 
mobility is very small. Thus, the bulk compositional perturbation 
$\hat{\phi}_b$ caused by the free surface disturbance is 
hypothesized to decay as
\begin{equation}
\hat{\phi}_b=\hat{\phi}_s e^{-\kappa(h_0-z)},
\label{eq-phi_b}
\end{equation}
with the corresponding $\hat{W}=\hat{\phi}_s e^{-\kappa(h_0-z)}
/(\kappa^2-q^2)$. This exponential form is for the early 
evolution regime and similar to that used in previous work.
\cite{glas,leonard98} The parameter $\kappa$ in Eq. 
(\ref{eq-phi_b}) is equal to $1/b$, with $b$ the vertical
length scale of compositional perturbation caused by free
surface. Due to the negligible atomic mobility in the bulk
and guided by the fact that the vertical morphological 
perturbation is very small compared with the lateral variation, 
one can assume $b\ll \lambda$, where $\lambda\sim 1/q$ is the
typical lateral wavelength of surface modulation. Therefore, 
for the range of $q$ that corresponds to typical surface
structure, we have
\begin{equation}
\kappa \gg q
\label{eq-kappa-q}
\end{equation}
in Eq. (\ref{eq-phi_b}).

Using the solutions of mechanical equilibrium equation and 
the assumptions (\ref{eq-phi_b}) and (\ref{eq-kappa-q}), we
have derived the dynamical equations for $\hat{h}$ and 
$\hat{\phi}_s$. To first order of the perturbations, they are
(in nondimensional form)
\begin{equation}
\partial\hat{h}^*/\partial \tau=({\epsilon^*}^2k^3-\gamma^*k^4)
\hat{h}^* - \frac{k^2}{1+\nu_0}\left [ \epsilon^*\eta^* + 
\frac{2E_1^*-(1+\nu_0)\mu_1^*}{2(1-\nu_0)}{\epsilon^*}^2 \right ]
\hat{\phi}_s^*,
\label{eq-h1}
\end{equation}
and
\begin{eqnarray}
\partial\hat{\phi}_s^*/\partial \tau&=&\frac{k^3}{1-\nu_0}\left [
(1-2\nu_0)\epsilon^*\eta^* + (2E_1^*-(1+\nu_0)\mu_1^*){\epsilon^*}^2
\right ]\hat{h}^* \nonumber\\
&-& k^2 \left [ k^2\pm 1+\frac{8E_1^*-5(1+\nu_0)\mu_1^*}
{2(1-\nu_0^2)}\epsilon^*\eta^* \right ]\hat{\phi}_s^*,
\label{eq-phi1}
\end{eqnarray}
where the ``$\pm$'' sign corresponds to the cases in which the alloy 
is above (``$+$'') or below (``$-$'') the effective critical
temperature $T_c^{\rm eff}$ defined by
\begin{equation}
T_c^{\rm eff}=T_c - \frac{2E_0}{1-\nu_0}\frac{\eta^2}{k_BN_v},
\label{eq-Tceff}
\end{equation}
which is the same as the spinodal temperature of alloys with 
coherency stress derived by Cahn.\cite{cahn}

Here we have rescaled the variables and parameters to make the
equations nondimensional, using a characteristic length scale
\begin{equation}
l_0 = \left ( \frac{|r|}{c}\right ) ^{-1/2},
\label{eq-l0}
\end{equation}
which is the typical width of domain interfaces, and time scale
\begin{equation}
\tau_0 = \left (\Gamma_{\phi}\frac{r^2}{c}\right )^{-1},
\label{eq-tau0}
\end{equation}
with 
\begin{equation}
r=r'-2E_0\eta^2/(1-\nu_0)=k_B(T_c^{\rm eff}-T)N_v,
\label{eq-r}
\end{equation}
as well as the transformations:
\begin{eqnarray}
k &=& ql_0, \nonumber\\
\tau &=& t/\tau_0, \nonumber\\
\hat{h}^* &=& h/l_0, \nonumber\\
\hat{\phi}_s^* &=& \hat{\phi}_s, \nonumber\\
\gamma^* &=& \frac{l_0}{c}\gamma, \nonumber\\
\epsilon^* &=& \left [\frac{2E_0}{|r|}\left (\frac{1+\nu_0}
{1-\nu_0}\right ) \right ]^{1/2} \epsilon, \nonumber\\
\eta^* &=& \left [\frac{2E_0}{|r|}\left (\frac{1+\nu_0}
{1-\nu_0}\right ) \right ]^{1/2} \eta.
\label{eq-scale}
\end{eqnarray}

In the early time regime of the perturbation's evolution, the
growth rates of morphological and compositional perturbations,
i.e., $\sigma_h$ and $\sigma_{\phi}$, are defined through
$\hat{h}^*=\hat{h}_0\exp(\sigma_h\tau)$ and $\hat{\phi}_s^*=
\hat{\phi}_0\exp(\sigma_{\phi}\tau)$, respectively. In general
cases (e.g., both $\epsilon^*$ and $\eta^*$ are nonzero) two
dynamical equations (\ref{eq-h1}) and (\ref{eq-phi1}) couple
with each other, which leads to the joint stability or 
instability of surface morphology and composition, that is,
$\sigma_h=\sigma_{\phi}=\sigma$. Therefore, from Eqs. 
(\ref{eq-h1}) and (\ref{eq-phi1}) we can obtain the 
characteristic equation for the perturbation growth rate 
$\sigma$:
\begin{eqnarray}
&&(\sigma+\gamma^*k^4-{\epsilon^*}^2 k^3)\left [ \sigma
+k^2\left (k^2\pm1+\frac{8E_1^*-5(1+\nu_0)\mu_1^*}{2(1-\nu_0^2)}
\epsilon^*\eta^* \right ) \right ] \nonumber\\
&&+\frac{k^5}{2(1+\nu_0)(1-\nu_0)^2}\left [ (1-2\nu_0)
\epsilon^*\eta^* + (2E_1^*-(1+\nu_0)\mu_1^*){\epsilon^*}^2
\right ] \nonumber\\
&&\times\left [ 2(1-\nu_0)\epsilon^*\eta^* + 
(2E_1^*-(1+\nu_0)\mu_1^*){\epsilon^*}^2 \right ] =0,
\label{eq-sigma}
\end{eqnarray}
where the ``$+$'' (``$-$'') sign corresponds to $T>T_c^{\rm eff}$
($T<T_c^{\rm eff}$). The real part of $\sigma$ determines the 
stability properties of the system: For Re$(\sigma)>0$, the film 
is jointly unstable to morphological instability and alloy  
decomposition instability at the surface, while if Re$(\sigma)<0$ 
is fulfilled for all the wavenumbers $k$, both the morphological 
and compositional modulations are suppressed. When Re$(\sigma)>0$ 
and Im$(\sigma)\neq 0$, the instability is oscillatory. In the 
following calculations, we first (Sec. \ref{secIV-A} and 
\ref{secIV-B}) focus on the real part of $\sigma$ for each of 
the solutions of Eq. (\ref{eq-sigma}), and use the largest one 
(with respect to all evolution modes $k$) to determine the regions
of instability. Within the unstable region, the imaginary part is 
then computed (Sec. \ref{secIV-C}) to determine the regions of 
oscillatory instability.

Note that in Eq. (\ref{eq-sigma}), quantities $k$, $\gamma^*$, 
$\epsilon^*$, $\eta^*$ depend on temperature via the coefficient 
$r$ (see Eqs.  (\ref{eq-r}), (\ref{eq-l0}), and (\ref{eq-scale})). 
The temperature dependence of the system's stability can be made 
explicit by rewriting Eq. (\ref{eq-sigma}) in a dimension-full 
manner, which is
\begin{eqnarray}
&&\left [ \Omega \tau_0 + \frac{c^{3/2}}
{|r|^{5/2}} \left ( \gamma q^4 - 2E_0 \frac{1+\nu_0}{1-\nu_0}
\epsilon^2 q^3 \right ) \right ] \nonumber\\
& \times& \left [ \Omega \tau_0 + \frac{c}{r^2}
q^2 \left ( -r + cq^2 + \frac{E_0}{(1-\nu_0)^2} (8E_1^* - 
5(1+\nu_0) \nu_1^* ) \epsilon \eta \right ) \right ] \nonumber\\
& +& \frac{c^{5/2}}{|r|^{9/2}} ~ \frac{2E_0^2(1+\nu_0)}{(1-\nu_0)^4}
~q^5 \left [ (1-2\nu_0)\epsilon\eta + (2E_1^*-(1+\nu_0)\mu_1^*)
\epsilon^2 \right ] \nonumber\\
& \times& \left [ 2(1-\nu_0)\epsilon\eta + 
(2E_1^*-(1+\nu_0)\mu_1^*)\epsilon^2 \right ] =0.
\label{eq-Omega}
\end{eqnarray}
Thus, in this characteristic equation for the perturbation growth 
rate $\sigma$ ($=\Omega \tau_0$), the temperature dependence enters 
only through the quantity $r$ which is linearly dependent on $T$ as 
given by Eq. (\ref{eq-r}). Even though the characteristic time scale
$\tau_0$ depends on temperature (Eq. (\ref{eq-tau0})), this 
dependence does not have any effect on stability boundaries.

\section{Results}
\label{secIV}

From Eqs. (\ref{eq-h1})--(\ref{eq-sigma}), the stability property
of the strained film depends on material parameters $\epsilon^*$,
$\eta^*$, $E_1^*$, $\mu_1^*$, $\gamma^*$, and $\nu_0$. In the
following we give the results of film stability for the cases of
composition independent and dependent elastic constants, as well
as different conditions of misfit and compositional strains.

\subsection{Composition-independent elastic moduli 
($E_1^*=\mu_1^*=0$)}
\label{secIV-A}

When ignoring the dependence of elastic constants on the local
composition, we have $E=E_0$, $\mu=\mu_0$, and $\nu=\nu_0$ from
Eq. (\ref{eq-Eu}), and the derived evolution equations for 
perturbations $\hat{h}^*$ and $\hat{\phi}_s^*$ as well as the 
characteristic equation for $\sigma$ are the same as Eqs. 
(\ref{eq-h1}), (\ref{eq-phi1}) and (\ref{eq-sigma}) after 
setting $E_1^*=\mu_1^*=0$.

When $\epsilon^*=\eta^*=0$, that is, neither misfit strain
nor compositional strain exists in the film, the dynamical 
equations for $\hat{h}^*$ and $\hat{\phi}_s^*$ decouple,
with different perturbation growth rate:
\begin{eqnarray}
\sigma_h &=& -\gamma^* k^4, \nonumber\\
\sigma_{\phi} &=& \left \{ \begin{array}{l}
-k^4 - k^2, \quad \textrm{if}~~ T>T_c^{\rm eff} \\
-k^4 + k^2, \quad \textrm{if}~~ T<T_c^{\rm eff}
\end{array} \right.
\label{eq-A1}
\end{eqnarray}
which recovers the results obtained by L\'{e}onard and Desai.
\cite{leonard98} Thus, the surface morphology is always stable 
and the compositional stability is similar to that of bulk alloy: 
For $T>T_c^{\rm eff}$ ($=T_c$, when $\eta=0$) the system is 
stable, while for $T<T_c^{\rm eff}$ spinodal decomposition 
occurs for long wavelengths ($k<1$). For the case of zero 
misfit but nonzero solute expansion coefficient, i.e., 
$\epsilon^*=0$ and $\eta^*\neq 0$, the dispersion relations 
are the same as (\ref{eq-A1}). Note that the compositional 
perturbation rate $\sigma_{\phi}$ obtained here is different 
from that of the previous work (see Eq. (44) in Ref. 24).

When $\epsilon^*\neq 0$ and $\eta^*=0$, corresponding to
nonzero misfit stress but zero solute stress, our results
yield:
\begin{eqnarray}
\sigma_h &=& {\epsilon^*}^2 k^3 - \gamma^* k^4, \nonumber\\
\sigma_{\phi} &=& \left \{ \begin{array}{l}
-k^4 - k^2, \quad \textrm{if}~~ T>T_c \\
-k^4 + k^2, \quad \textrm{if}~~ T<T_c
\end{array} \right.
\label{eq-A2}
\end{eqnarray}
where the Asaro-Tiller-Grinfeld instability for morphology 
\cite{asaro,grinfeld} is recovered, as also shown in the
model of L\'{e}onard and Desai.

For more general case of $\epsilon^*\neq 0$ and $\eta^*\neq 0$,
that is, the strains generated by both the lattice mismatch and
compositional nonuniformity are nonzero and coupled, we have
a quadratic equation for the common perturbation growth rate
$\sigma$:
\begin{equation}
\sigma^2 + a_1 \sigma + a_0 =0,
\label{eq-sigmaA}
\end{equation}
with the roots $\sigma=\left (-a_1\pm\sqrt{a_1^2-4a_0}\right )/2$ 
and coefficients
\begin{eqnarray}
a_1 &=& \gamma^* k^4 - {\epsilon^*}^2 k^3 + k^2(k^2\pm 1), 
\nonumber\\
a_0 &=& k^2(k^2\pm 1)(\gamma^* k^4 - {\epsilon^*}^2 k^3)
+ \frac{1-2\nu_0}{1-\nu_0^2}{\epsilon^*}^2{\eta^*}^2 k^5.
\label{eq-aA}
\end{eqnarray}
Usually the Poisson ratio $\nu_0$ lies in the range from $1/4$ to
$1/3$. Consequently, for $T<T_c^{\rm eff}$ (bottom sign ``$-$'' 
in Eq. (\ref{eq-aA})), we have $a_1 <0$ and $a_0 >0$ when the 
wavenumber is very small, i.e., $k\ll 1$, corresponding to the 
solution Re$(\sigma)>0$. Then the instabilities of surface
morphological and compositional profiles are expected to appear
simultaneously below the effective critical temperature.

For $T>T_c^{\rm eff}$, the stability properties are more
complicated and we present the analytic results as follows.
If ${\gamma^*}^2 > \gamma^*+1$ (i.e., $\gamma^* > (1+\sqrt{5})/2$), 
the stability condition for this strained film is
\begin{equation}
{\epsilon^*}^2 < 2(1+\gamma^*)^{1/2} \qquad {\rm and} \qquad
{\eta^*}^2 > {\eta_0^*}^2,
\label{eq-stab1A}
\end{equation}
with ${\eta_0^*}^2=(1-\nu_0^2)/(1-2\nu_0)$. Otherwise, if
${\gamma^*}^2 < \gamma^*+1$ (i.e., $0 < \gamma^* < (1+\sqrt{5})/2$), 
the system is stable when
\begin{equation}
{\epsilon^*}^2 < 2\gamma^* \qquad {\rm and} \qquad
{\eta^*}^2 > {\eta_0^*}^2,
\label{eq-stab2A}
\end{equation}
or
\begin{equation}
2\gamma^* < {\epsilon^*}^2 < 2(1+\gamma^*)^{1/2} 
\quad {\rm and} \quad {\eta^*}^2 > {\eta_0^*}^2
\left [ 1-\frac{{\epsilon^*}^2(9{\gamma^*}^2-2{\epsilon^*}^4)
- 2({\epsilon^*}^4-3{\gamma^*}^2)^{3/2}}
{27{\gamma^*}^2 {\epsilon^*}^2} \right ].
\label{eq-stab3A}
\end{equation}
Therefore, in this stress-driven epitaxial system the instability
of both morphology and composition could also occur above the
strained spinodal temperature $T_c^{\rm eff}$ and for large misfit
$\epsilon^*$ or small solute coefficient $\eta^*$. This result is 
very different from the bulk alloy where only below the critical 
temperature, can the spinodal decomposition be present. This 
instability is due to the coupling of morphological and compositional 
undulations, as pointed out by Glas \cite{glas} from the thermodynamic 
point of view.

The corresponding stability boundary in the $\epsilon^*$--$\eta^*$ 
space is shown for $\nu_0=1/4$ and different values of $\gamma^*$
in Fig. \ref{fig-A+} (thick lines). The stability boundary for 
$\gamma^*=5$ corresponds to the case of ${\gamma^*}^2 > \gamma^*
+1$ and then is determined by Eq. (\ref{eq-stab1A}), while for 
$\gamma^*=0.5$ the conditions (\ref{eq-stab2A}) and (\ref{eq-stab3A}) 
are used. The stabilizing effect of surface energy can be seen 
directly from the figure, where the stable region is enlarged with 
the increasing value of surface tension $\gamma^*$. The stability
diagram here is symmetric with respect to the sign of misfit
$\epsilon^*$ since we have assumed the composition independence
of elastic moduli in Fig. \ref{fig-A+}.

\subsection{Composition-dependent elastic moduli 
($E_1^*\neq 0$, $\mu_1^*\neq 0$)}
\label{secIV-B}

More interesting and richer results are obtained for the
cases of composition-dependent elastic constants with 
$E_1^*\neq 0$ and $\mu_1^*\neq 0$, where the coupling of
misfit strain, solute strain and composition dependence of
elastic moduli can highly affect the behaviors of perturbation
growth.

For the lattice matched films, that is, $\epsilon^*=0$ with
$\eta^*$ arbitrary, the morphological and compositional 
degrees of freedom decouple, as obtained from dynamical
equations (\ref{eq-h1}) and (\ref{eq-phi1}). The
perturbation growth rates $\sigma_h$ and $\sigma_{\phi}$
also obey Eq. (\ref{eq-A1}), corresponding to the stability
properties the same as that of $E_1^*=\mu_1^*=0$. However,
when lattice misfit exists, i.e., $\epsilon^*\neq 0$,
the composition dependence of elastic constants leads to
substantially different results. In the absence of atomic
size difference ($\eta^*=0$), the dynamical equations for
$\hat{h}^*$ and $\hat{\phi}_s^*$ remain coupled, which is
qualitatively different from the case shown in Eq. 
(\ref{eq-A2}) for composition independent elastic moduli. 
The coupled perturbation growth rate $\sigma$ is then
governed by a quadratic equation of the form similar to
Eq. (\ref{eq-sigmaA})
$$\sigma^2 + a_1 \sigma + a_0 =0,$$
but with different coefficient $a_0$:
\begin{eqnarray}
a_1 &=& \gamma^* k^4 - {\epsilon^*}^2 k^3 + k^2(k^2\pm 1), 
\nonumber\\
a_0 &=& k^2(k^2\pm 1)(\gamma^* k^4 - {\epsilon^*}^2 k^3)
+\frac{[2E_1^*-(1+\nu_0)\mu_1^*]^2}{2(1+\nu_0)(1-\nu_0)^2}
{\epsilon^*}^4 k^5.
\label{eq-aB1}
\end{eqnarray}
When $T<T_c^{\rm eff}$, it is easy to show that 
Re$(\sigma)>0$ for $k\ll 1$ and then the system is unstable,
while for $T>T_c^{\rm eff}$ the film can be stable for 
certain values of misfit $\epsilon^*$, as specified in
the following stability conditions:
If ${\gamma^*}^2 > \gamma^*+1$, the stability occurs for
\begin{equation}
\chi^{-1} < {\epsilon^*}^2 < 2(1+\gamma^*)^{1/2},
\label{eq-stabB1_1}
\end{equation}
with
\begin{equation}
\chi=\frac{[2E_1^*-(1+\nu_0)\mu_1^*]^2}
{2(1+\nu_0)(1-\nu_0)^2};
\label{eq-chi}
\end{equation}
while for ${\gamma^*}^2 < \gamma^*+1$, there are two regions
of stability. In the first one,
\begin{equation}
\chi^{-1} < {\epsilon^*}^2 < 2\gamma^*,
\label{eq-stabB1_2}
\end{equation}
is fulfilled. In the other, four conditions have to be 
fulfilled:
\begin{eqnarray}
&& {\rm (i)\: } 2\gamma^* < {\epsilon^*}^2 < 2(1+\gamma^*)^{1/2}, \\
\label{eq-stabB1_31}
&& {\rm (ii)\: } \gamma^* > \frac{4}{9}\chi^{-1}, \\
\label{eq-stabB1_32}
&& {\rm (iii)\: } \gamma^* > \frac{{\epsilon^*}^2}
{3(3\chi{\epsilon^*}^2/2-1)^{1/2}},
\label{eq-stabB1_33}
\end{eqnarray}
and (iv)
\begin{eqnarray}
&& \frac{2}{3}\chi^{-1} < {\epsilon^*}^2 < 
\frac{8}{9}\chi^{-1}, \nonumber\\
&& {\rm or}  \nonumber\\
&& {\epsilon^*}^2 > \frac{8}{9}\chi^{-1}, \:\, {\rm and} \:\,
{\gamma^*}^2 > \frac{{\epsilon^*}^4}{2} \left [ \left( 
\frac{\chi}{2} \right )^{1/2} \left( \frac{9}{2}\chi{\epsilon^*}^2
-4 \right )^{3/2}\epsilon^* - 3\left ( \frac{3}{2}
\chi{\epsilon^*}^2-1 \right )^2+1 \right ], \nonumber\\
&& {\rm or} \nonumber\\
&& {\epsilon^*}^2 > \frac{8}{9}\chi^{-1}, \:\, {\rm and} \:\,
{\gamma^*}^2 < \frac{{\epsilon^*}^4}{2} \left [ -\left( 
\frac{\chi}{2} \right )^{1/2} \left( \frac{9}{2}\chi{\epsilon^*}^2
-4 \right )^{3/2}\epsilon^* - 3\left ( \frac{3}{2}
\chi{\epsilon^*}^2-1 \right )^2+1 \right ].
\label{eq-stabB1_34}
\end{eqnarray}
The stability diagram of $|\epsilon^*|$ vs $\gamma^*$ 
corresponding to Eqs. (\ref{eq-stabB1_1})--(\ref{eq-stabB1_33}) 
is plotted in Fig. \ref{fig-eta0+}, where the parameters 
$\nu_0=1/4$, $E_1^*=-0.4$, and $\mu_1^*=-0.1$ are chosen.
One can see from the diagram that in the absence of 
compositional strain but with the composition dependence of 
elastic constants, the system above the effective critical
temperature can be stabilized for intermediate magnitudes of 
misfit $\epsilon^*$ and large enough effective surface tension 
$\gamma^*$.

For the most general case $\epsilon^*\neq 0$, $\eta^*\neq 0$
and $E_1^*\neq 0$, $\mu_1^*\neq 0$, corresponding to the 
lattice mismatched and compositionally stressed film with 
composition-dependent elastic constants, the coupled dynamical 
equations are described in (\ref{eq-h1}) and (\ref{eq-phi1}), 
with joint perturbation growth rate $\sigma$ given by Eq. 
(\ref{eq-sigma}). The characteristic equation (\ref{eq-sigma}) 
is in fact quadratic, with coefficients
\begin{eqnarray}
a_1 &=& \gamma^* k^4 - {\epsilon^*}^2 k^3 + 
k^2\left [k^2\pm1+\beta\epsilon^*\eta^* \right ], \nonumber\\
a_0 &=& k^2(\gamma^* k^4 - {\epsilon^*}^2 k^3)
\left [k^2\pm1+\beta \epsilon^*\eta^* \right ] \nonumber\\
&+& \frac{k^5}{2(1+\nu_0)(1-\nu_0)^2}\left [ (1-2\nu_0)
\epsilon^*\eta^* + \alpha{\epsilon^*}^2
\right ] \left [ 2(1-\nu_0)\epsilon^*\eta^* + 
\alpha{\epsilon^*}^2 \right ],
\label{eq-aB2}
\end{eqnarray}
with the parameters
\begin{eqnarray}
\alpha &=& 2E_1^*-(1+\nu_0)\mu_1^*, \nonumber\\
\beta &=& \frac{8E_1^*-5(1+\nu_0)\mu_1^*}{2(1-\nu_0^2)}.
\label{eq-alpha-beta}
\end{eqnarray}
The stability conditions can be derived by studying the real part
of the solutions $\sigma=\left (-a_1\pm\sqrt{a_1^2-4a_0}\right )/2$,
and we present the analytic results below for both $T>T_c^{\rm eff}$
and $T<T_c^{\rm eff}$.

The stable epitaxial film should first fulfill:
\begin{equation}
\beta \epsilon^* \eta^* \pm 1 >0,
\label{eq-stabB2_1}
\end{equation}
and then similar to the other cases, the conditions for 
${\gamma^*}^2 > \gamma^*+1$ or ${\gamma^*}^2 < \gamma^*+1$ are 
different. For ${\gamma^*}^2 > \gamma^*+1$, the stability 
conditions are
\begin{equation}
{\epsilon^*}^4 < 4(\gamma^*+1) (\beta \epsilon^* \eta^* \pm 1),
\label{eq-stabB2_21}
\end{equation}
and
\begin{eqnarray}
&& \Delta <0, \nonumber\\
&&{\rm or} \nonumber\\
&& \Delta > 0 \quad {\rm and} \quad
 \eta^* > \frac{-\rho \epsilon^* + \Delta^{1/2}}
{4(1-\nu_0)(1-2\nu_0)}, \nonumber\\
&&{\rm or} \nonumber\\
&& \Delta > 0 \quad {\rm and} \quad
 \eta^* < \frac{-\rho \epsilon^* - \Delta^{1/2}}
{4(1-\nu_0)(1-2\nu_0)},
\label{eq-stabB2_22}
\end{eqnarray}
where
\begin{eqnarray}
\rho &=& -2E_1^* + (1+\nu_0)(2-\nu_0)\mu_1^*, \nonumber\\
\Delta &=& \left [ \rho^2
-8(1-\nu_0)(1-2\nu_0)\alpha^2 \right ]{\epsilon^*}^2
\pm 16(1+\nu_0)(1-\nu_0)^3(1-2\nu_0).
\label{eq-rho-Del}
\end{eqnarray}
On the other hand, for ${\gamma^*}^2<\gamma^*+1$ the system
is stable only when the following conditions are fulfilled:
\begin{equation}
{\epsilon^*}^4 < 4{\gamma^*}^2 (\beta \epsilon^* \eta^* \pm 1),
\label{eq-stabB2_31}
\end{equation}
as well as all the conditions in Eq. (\ref{eq-stabB2_22}), or
\begin{eqnarray}
&& 4{\gamma^*}^2 (\beta \epsilon^* \eta^* \pm 1) <{\epsilon^*}^4 
<4(\gamma^*+1) (\beta \epsilon^* \eta^* \pm 1), \nonumber\\
&& {\rm and} \nonumber\\
&& \left [ \frac{1-2\nu_0}{1-\nu_0^2}{\eta^*}^2 + \frac{\rho}
{2(1+\nu_0)(1-\nu_0)^2}\epsilon^*\eta^* +\frac{\alpha^2}
{2(1+\nu_0)(1-\nu_0)^2}{\epsilon^*}^2 \mp 1 \right ]
{\epsilon^*}^2 \nonumber\\
&& +\frac{{\epsilon^*}^2}{27{\gamma^*}^2}[9{\gamma^*}^2
(\beta \epsilon^* \eta^* \pm 1) - 2{\epsilon^*}^4]
- \frac{2}{27{\gamma^*}^2}[{\epsilon^*}^4-3{\gamma^*}^2
(\beta \epsilon^* \eta^* \pm 1)]^{3/2} >0.
\label{eq-stabB2_32}
\end{eqnarray}
Note that in Eqs. (\ref{eq-aB2}), (\ref{eq-stabB2_1}),
(\ref{eq-stabB2_21}), (\ref{eq-rho-Del})--(\ref{eq-stabB2_32}), 
the top sign applies when the temperature $T$ is above the 
effective critical temperature $T_c^{\rm eff}$, and the bottom
sign corresponds to $T<T_c^{\rm eff}$.

The stability diagrams can be calculated according to above 
results (\ref{eq-stabB2_1})--(\ref{eq-stabB2_32}). Here we
use two sets of parameters (1 and 2) to plot the stability 
diagrams of $T>T_c^{\rm eff}$ and $T<T_c^{\rm eff}$, as shown 
in Figs. \ref{fig-B1+}--\ref{fig-B2-}. For the first set
(Set 1, used in Figs. \ref{fig-B1+} and \ref{fig-B1-}), 
where all the material parameters (e.g., $T_c$, $\gamma$, 
$a_f$, $N_v$, and elastic constants) are chosen to 
qualitatively represent the SiGe alloy, we have $\nu_0=1/4$ 
and $\gamma^*=5$ obtained from Eq. (\ref{eq-scale}), and 
assume that $E_1^*=-0.4$, $\mu_1^*=-0.1$. The second set 2 
is expected to qualitatively represent the InGaAs alloy and 
applies to Figs. \ref{fig-B2+} and \ref{fig-B2-}, where we 
choose $\nu_0=1/3$, $\gamma^*=3.5$, $E_1^*=-0.25$, and
$\mu_1^*=-0.5$.

Compared to symmetric diagram Fig. \ref{fig-A+} for 
$E_1^*=\mu_1^*=0$, Figs. \ref{fig-B1+}--\ref{fig-B2-} exhibit 
the asymmetry for compressive and tensile layers, i.e., the
stability depends on the sign of misfit $\epsilon^*$, which 
is one of the major consequence of local composition dependence 
of elastic constants. In Figs. \ref{fig-B1+} and \ref{fig-B1-},
corresponding to parameters of Set 1 (similar to the case of
SiGe), the stabilization mainly occurs under tensile strain 
($\epsilon^*<0$) and the stable regions increase with larger 
value of $\eta^*$, while for compressive films the instability 
can not be suppressed for most of the parameter values, 
especially for $T<T_c^{\rm eff}$. In contrast, the InGaAs-like
parameters of Set 2 lead to opposite asymmetry: Larger part of
stable region is found in positive misfit $\epsilon^*$, and
layers subject to tensile strain exhibit less stability, as
shown in Figs. \ref{fig-B2+} and \ref{fig-B2-}.

The other important effect of composition-dependent elastic
moduli is on the system below effective critical temperature
$T_c^{\rm eff}$. For all the other cases described above,
including the ones for $E_1^*\neq 0$ and $\mu_1^*\neq 0$ but
with one of $\epsilon^*$ and $\eta^*$ equal to zero, the
compositional profiles for $T<T_c^{\rm eff}$ are unstable,
in agreement with the usual expectation that the strained
alloy near $50-50$ mixture should exhibit decomposition
and phase segregation below the coherent spinodal temperature 
$T_c^{\rm eff}$. However, the coupling of all the factors of 
misfit strain, solute strain and composition-dependent moduli 
causes different and new effects. As shown in Figs. 
\ref{fig-B1-} and \ref{fig-B2-} when all the variables 
$\epsilon^*$, $\eta^*$, $E_1^*$ and $\mu_1^*$ are nonzero, 
the film below $T_c^{\rm eff}$ can also be stable for certain 
range of parameters. That is, it is possible to suppress the 
surface decomposition even for $T<T_c^{\rm eff}$ due to the 
coupling effects in this heteroepitaxial system.

\subsection{Oscillatory instability (Im$(\sigma)\neq 0$)}
\label{secIV-C}

When the system corresponds to the unstable parameters 
region of stability diagram, that is, Re$(\sigma)>0$, the 
imaginary part of $\sigma$ determines whether the onset of 
this instability is steady (Im$(\sigma)=0$) or oscillatory
(Im$(\sigma)\neq 0$). The occurrence of oscillatory
instability has been found in the study of directional
solidification for stressed solid \cite{spencer92} and 
the growing process of alloy thin films, \cite{guyer96}
and been attributed to the phase difference between surface
morphology and composition field, induced by nonlocal
elastic stresses.

We calculate the imaginary part of perturbation growth rate
$\sigma$ through characteristic equation (\ref{eq-sigma}).
If for a certain wavenumber $k$, we have both Re$(\sigma)>0$
and Im$(\sigma)\neq 0$, the oscillatory instability may occur.
The results for $\epsilon^* \neq 0$, $\eta^* \neq 0$, and
$T>T_c^{\rm eff}$, corresponding to the parameters range of
most experiments on strained films, are shown in Figs. 
\ref{fig-A+}, \ref{fig-B1+}, and \ref{fig-B2+} with the thin 
dashed or dotted boundary curves. For the case of composition
independent elastic moduli, as shown in Fig. \ref{fig-A+}, 
the oscillatory unstable regions are symmetric with respect 
to misfit and more regular. Large solute coefficient $\eta^*$ 
favours the occurrence of oscillatory instability, and if 
misfit $\epsilon^*$ is large enough, the oscillatory instability
is obtained when $\eta^*$ exceeds a fixed value $\eta^*_0 =
[(1-\nu_0^2)/(1-2\nu_0)]^{1/2}$, which can be derived
analytically. When the elastic constants are composition 
dependent, \textit{i.e.} $E_1^*\neq 0$ and $\mu_1^*\neq 0$,
the oscillatory regions are asymmetric and irregular (see
Figs. \ref{fig-B1+} and \ref{fig-B2+}). Note that in these 
parameter regions of oscillatory instability, steady instability 
can also exist (but for different mode $k$), and the competition 
between these two kinds of unstable modes determines the surface 
profile. When oscillatory modes dominate or coexist with steady 
modes, the surface disturbance is expected to propagate laterally, 
with the phenomenon that one side of the surface bump will grow 
faster than the other side. \cite{spencer92}

\section{Discussion and conclusions}
\label{secV}

\begin{figure}
\centerline{
\resizebox{0.5\textwidth}{!}{%
  \includegraphics{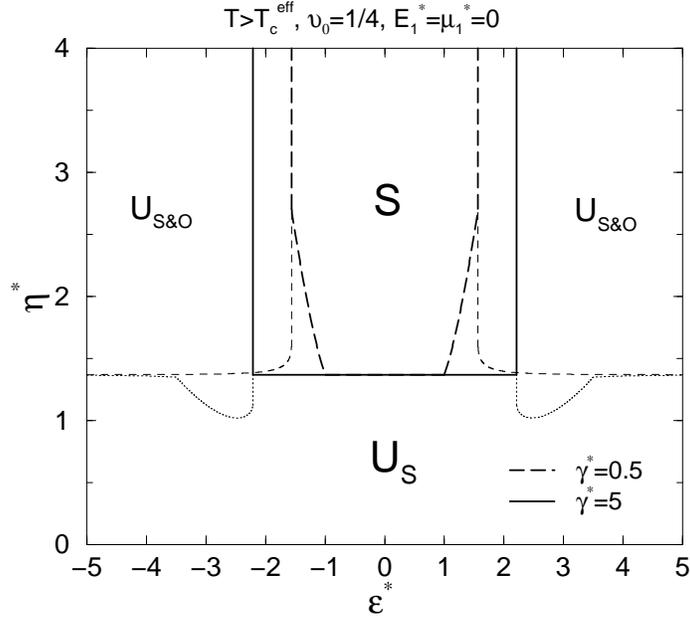}}}
\caption{Stability diagram for non-growing strained alloy film
with $E_1^*=\mu_1^*=0$, temperature $T>T_c^{\rm eff}$, and 
$\nu_0=1/4$. Region marked by ``S'' is the stable region. Long 
dashed and solid thick lines denote the stability boundaries 
for $\gamma^*=0.5$ and $5$, respectively. In the unstable region, 
the domain marked as ``U$_{\textrm S}$'' corresponds to steady 
instability and that marked as ``U$_\textrm{S\&O}$'' corresponds 
to steady or oscillatory instabilities depending on the 
wavenumber $k$. The boundaries between ``U$_{\textrm S}$'' 
and ``U$_\textrm{S\&O}$'' regions are indicated by dashed 
($\gamma^*=0.5$) and dotted ($\gamma^*=5$) thin curves.}
\label{fig-A+}
\end{figure}

\begin{figure}
\centerline{
\resizebox{0.5\textwidth}{!}{%
  \includegraphics{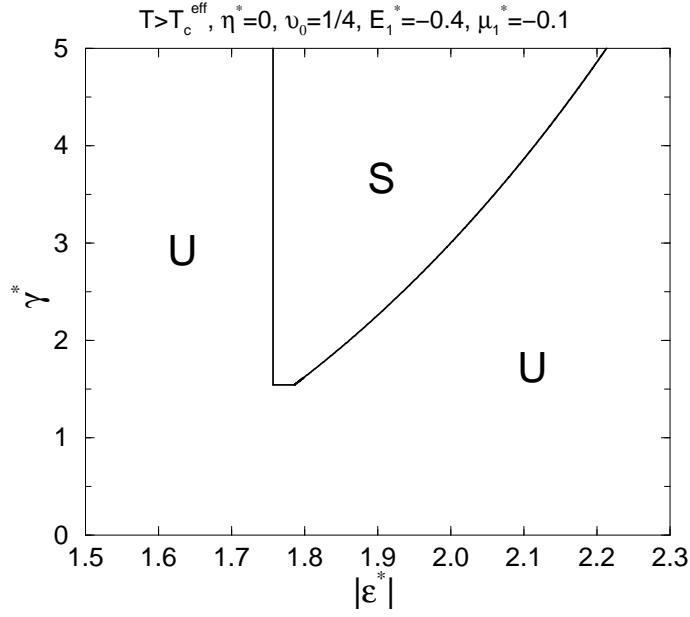}}}
\caption{Stability diagram for $\eta^*=0$ non-growing strained 
alloy film, with $T>T_c^{\rm eff}$, $\nu_0=1/4$, $E_1^*=-0.4$, 
and $\mu_1^*=-0.1$. Stable and unstable regions are marked as 
``S'' and ``U'', respectively.}
\label{fig-eta0+}
\end{figure}

\begin{figure}
\centerline{
\resizebox{0.53\textwidth}{!}{%
  \includegraphics{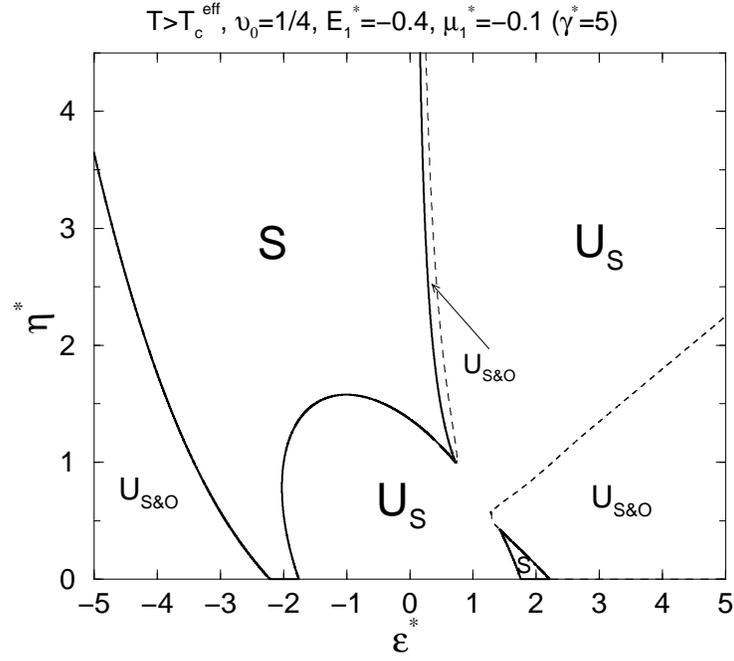}}}
\caption{Stability diagram for non-growing strained alloy 
film with $T>T_c^{\rm eff}$. Parameters of Set 1 are chosen: 
$E_1^*=-0.4$, $\mu_1^*=-0.1$, $\nu_0=1/4$, and $\gamma^*=5$.
Stable and unstable regions are marked in a manner similar
to that in Fig. \ref{fig-A+}.}
\label{fig-B1+}
\end{figure}

\begin{figure}
\centerline{
\resizebox{0.53\textwidth}{!}{%
  \includegraphics{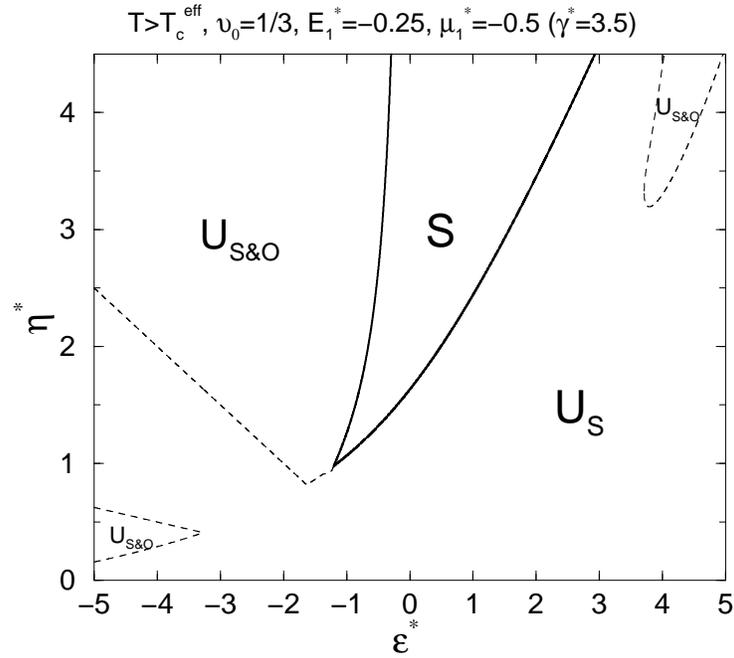}}}
\caption{Stability diagram for $T>T_c^{\rm eff}$ strained alloy 
film in the absence of growth, with Set 2 parameters: $E_1^*=-0.25$, 
$\mu_1^*=-0.5$, $\nu_0=1/3$, and $\gamma^*=3.5$. Regions with 
different stability properties are marked as in Figs. \ref{fig-A+}
and \ref{fig-B1+}.}
\label{fig-B2+}
\end{figure}

\begin{figure}
\centerline{
\resizebox{0.53\textwidth}{!}{%
  \includegraphics{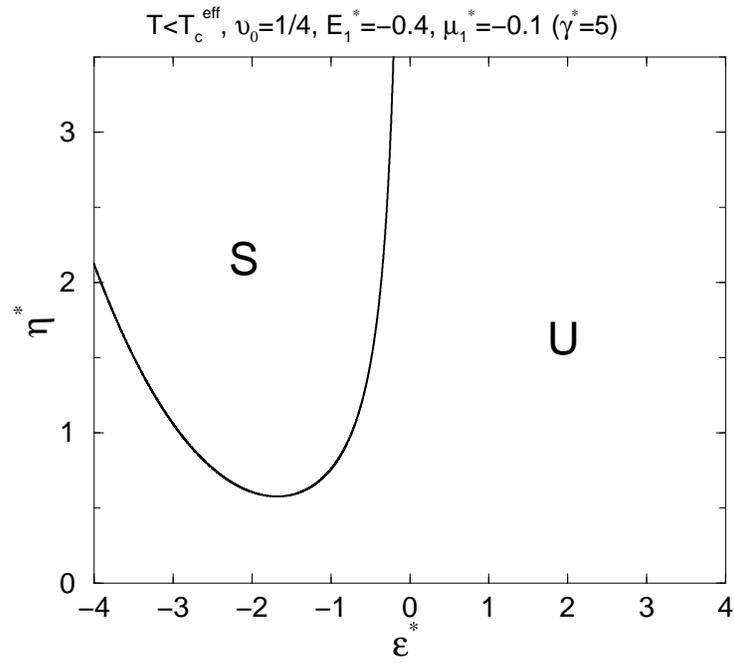}}}
\caption{Stability diagram for non-growing strained alloy film
with $T<T_c^{\rm eff}$ and composition-dependent elastic moduli. 
The parameters are of Set 1 as described in Fig. \ref{fig-B1+}.}
\label{fig-B1-}
\end{figure}

\begin{figure}
\centerline{
\resizebox{0.53\textwidth}{!}{%
  \includegraphics{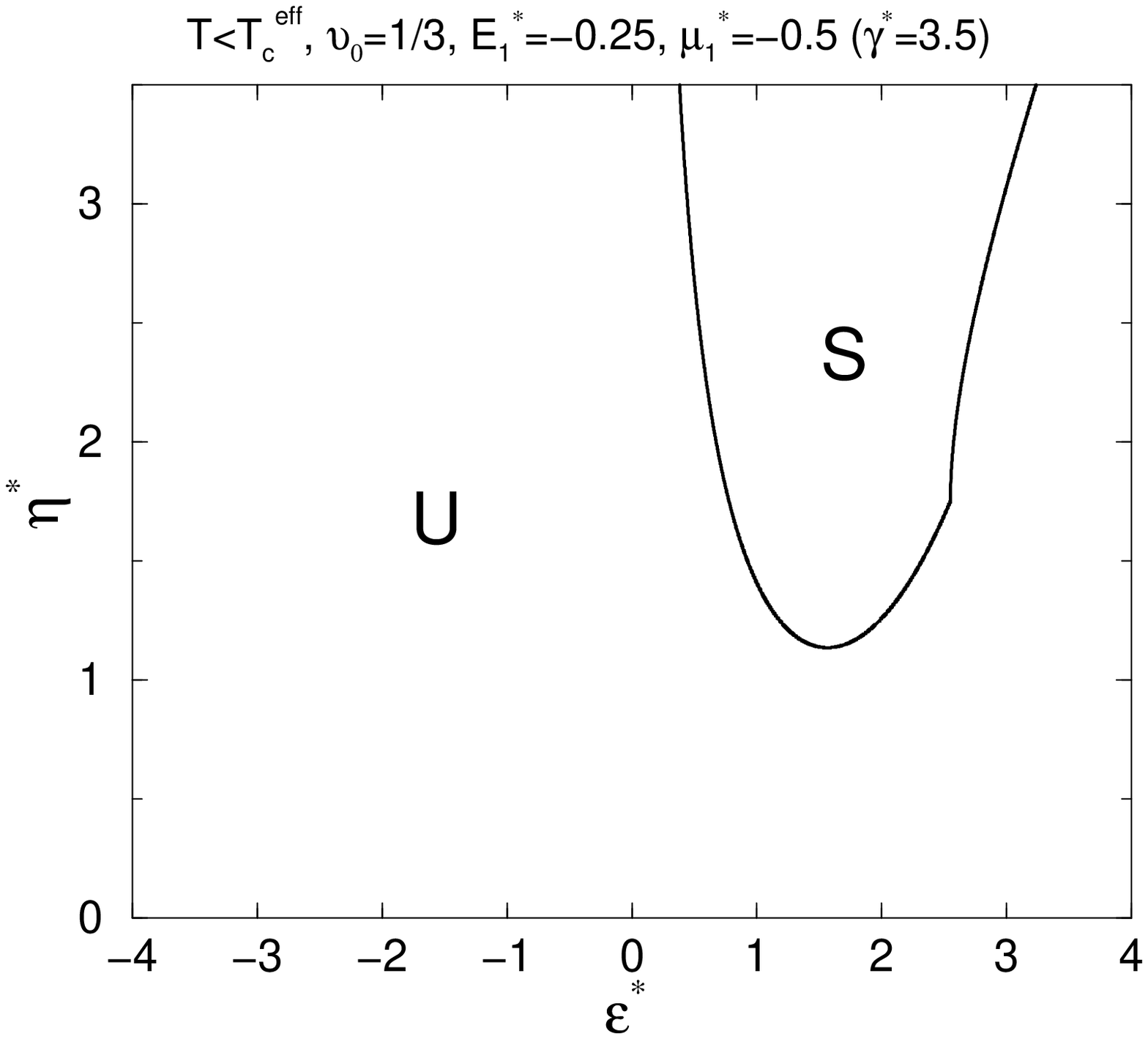}}}
\caption{Stability diagram for non-growing strained alloy film
with $T<T_c^{\rm eff}$ as well as the Set 2 parameters as
described in Fig. \ref{fig-B2+}.}
\label{fig-B2-}
\end{figure}

Our calculations above have shown that the stability problem
of free alloy film surface under both morphological and 
compositional strains is essentially a nonequilibrium and
dynamical problem even for static films, i.e., in the absence 
of growth. Although in the sense of thermodynamics and 
equilibrium, it has been demonstrated that the instability 
can appear in any stressed alloy with free surface,\cite{glas} 
the physically based choice of non-equilibrium evolution 
dynamics leads to different conclusion that the system 
can be stabilized for certain values of parameters. As shown 
in Figs. \ref{fig-A+}--\ref{fig-B2-}, the joint instability
can be suppressed by large enough compositional strain
($\eta^*$), which has also been found for growing films
in the film-vapor local equilibrium model of Guyer and
Voorhees \cite{guyer96} as well as in the dynamical model
of Spencer et al. \cite{spencer01} with unequal atomic 
mobilities for different alloy constituents.

In the previous work \cite{guyer96,spencer01,leonard98} this 
stabilization of compositional stresses would be overcome by 
larger magnitude of misfit $\epsilon^*$, while in our results 
similar phenomenon occurs for $\epsilon^*$ values far from 
zero, but the other parts of stability diagrams are more 
complicated. In Fig. \ref{fig-B2+} (with parameters of Set 
2 and $T>T_c^{\rm eff}$) and Figs. \ref{fig-B1-} and 
\ref{fig-B2-} for $T<T_c^{\rm eff}$, when the magnitude of
misfit $\epsilon^*$ is close to the minimum of the stability
boundary and is made smaller, then higher value of $\eta^*$ 
(related to larger compositional strain) is needed to suppress 
the instability, and for parameters of Set 1 with 
$T>T_c^{\rm eff}$ (Fig. \ref{fig-B1+}) the stable regions 
are much more irregular, due to the combination of misfit 
and compositional strains as well as composition-dependent 
elastic moduli. Even for the case of composition-independent 
moduli, the results here (shown in Fig. \ref{fig-A+}) are 
different from before. For sufficiently large $\eta$, the 
nonphysical short wavelength divergence \cite{guyer96,
spencer01} is avoided due to the inclusion of gradient energy 
and the return of instability \cite{leonard98} is not found 
here due to the consideration of coupling between surface and 
bulk phases.

The introduction of composition dependence of alloy elastic 
constants, which makes the effective elastic effects nonlocal, 
leads to the presence of asymmetry in the stability of films 
under compression or tension. Some experiments have tested this 
misfit sign dependence. Although these observations are all 
for the growing films, they can still be helpful to check our 
theoretical results for films without deposition. Our 
calculations using parameters similar to that of SiGe (shown 
in Figs. \ref{fig-B1+} and \ref{fig-B1-}) exhibit the preference 
of stability for tensile layers, consistent with the experimental
findings of Xie \textit{et al.}. \cite{xie} Various experiments
\cite{gendry,guyer00,okada} indicated that the way of asymmetry 
for InGaAs alloys depends on the materials deposition rate, 
while our results in Figs. \ref{fig-B2+} and \ref{fig-B2-} with 
a specific selection of parameters (Set 2) suggest that the 
compressive layers are more stable for non-growing film.

The simultaneous interaction of misfit strain, solute strain 
and composition-dependent elastic constants makes the stability
possible even for $T<T_c^{\rm eff}$, which is not possible
in the absence of any one of them. The corresponding diagrams 
are shown in Figs. \ref{fig-B1-} and \ref{fig-B2-}. This 
phenomenon is very different from what we expect, since in usual 
bulk strained alloy the spinodal decomposition always occurs 
below $T_c^{\rm eff}$.\cite{cahn} The related experiments are 
lacking since most of the epitaxial experiments are carried 
out above the effective critical temperature $T_c^{\rm eff}$.

Note that although in the above analysis, we have distinguished 
the stability results and diagrams into two temperature regimes 
$T>T_c^{\rm eff}$ and $T<T_c^{\rm eff}$, in each regime the 
stability properties are still temperature dependent. This can
be seen from our theoretical diagrams (Figs. \ref{fig-A+} -- 
\ref{fig-B2-}) as well as the charateristic equation 
(\ref{eq-sigma}) for perturbation growth rate, where the system 
stability is shown to depend on the rescaled parameters 
$\epsilon^*$, $\eta^*$, and $\gamma^*$, which in fact are all 
proportional to $|r|^{-1/2}$ with $r$ linearly dependent on 
temperature $T$ (see Eqs. (\ref{eq-scale}) and (\ref{eq-r})).
This temperature dependence of stability property can also be
obtained from Eq. (\ref{eq-Omega}) that has dimension-full form.

It is interesting to compare our theoretical results for 
instability with the observations of isothermal annealing 
experiments on strained films. Experiments on SiGe/Si 
postdeposition system \cite{walther,chen,ozkan} have exhibited 
a morphological evolution procedure from an initial planar 
film surface to a rough surface profile with ripples or islands 
during annealing. Also, along with this morphological modulation, 
the Ge segregation has been found on the surface of SiGe layers 
by Walther \textit{et al.}, \cite{walther} which corresponds to 
the coupling of compositional and morphological instabilities 
studied here. The temperature effect on the stability of 
Si$_{0.5}$Ge$_{0.5}$ strained films has been investigated by 
Chen \textit{et al.}. \cite{chen} They observed that the 
evolution to unstable surface morphology only occurs above
an annealing temperature, and attributed this sharp temperature
dependence to the existence of an energy barrier. Our theoretical 
results (for the case of $T>T_c^{\rm eff}$, e.g., stability 
diagram like Fig. \ref{fig-B1+}), obtained from the surface 
diffusion mechanism, can explain this temperature dependent 
phenomenon without the introduction of an energy barrier. For
small temperature $T$ (but still above $T_c^{\rm eff}$, as in
real experiments), we have small $|r|$ (see Eq. (\ref{eq-r})) 
and then large $\gamma^*$ (resulting in the large stable region 
of the stability diagram) and large $\eta^*$, which are apt to 
suppress the instability. When $T$ increases, the value of 
$\epsilon^*$ becomes smaller, and more importantly, $\gamma^*$ 
and $\eta^*$ also decrease, rendering the system closer to the 
unstable region of the stability diagram. Thus, one can reach a 
transition temperature above which the annealing system is 
unstable, as found in the experiments.

In summary, we have developed a continuum model to study the
nonequilibrium evolution processes of strained alloy films
in the absence of growth. With the consideration of coupling
between surface and bulk states as well as the composition
dependence of elastic moduli (both $E$ and $\mu$), new and
more complicated stability results and diagrams have been 
obtained using linear stability analysis. In general case, 
joint stability or instability is found due to the coupling 
between morphological and compositional perturbations. More 
importantly, the interplay of morphological and compositional 
strains as well as the composition-dependent elastic constants 
leads to the stability dependence on the sign of film-substrate 
misfit, and the possibility of stabilization even for films 
below $T_c^{\rm eff}$. Here we only study the early film 
evolution, and the nonlinear effects should be considered for 
later regimes and for the determination of detailed surface 
morphologies and patterns.

\section*{Acknowledgments}

This work was supported by the NSERC of Canada.

\newpage

\begin{thebibliography}{}

\bibitem{asaro} R. J. Asaro and W. A. Tiller, Metall. Trans. {\bf 3}, 
1789 (1972).

\bibitem{grinfeld} M. A. Grinfeld, Sov. Phys. Dokl. {\bf 31}, 831 (1987).

\bibitem{srolovitz} D. J. Srolovitz, Acta Metall. {\bf 37}, 621 (1989).

\bibitem{spencer91} B. J. Spencer, P. W. Voorhees, and S. H. Davis, Phys.
Rev. Lett. {\bf 67}, 3696 (1991); J. Appl. Phys. {\bf 73}, 4955 (1993).

\bibitem{legoues} F. K. LeGoues, M. Copel, and R. M. Tromp, Phys. Rev. B
{\bf 42}, 11690 (1990).

\bibitem{eaglesham} D. J. Eaglesham and M. Cerullo, Phys. Rev. Lett.
{\bf 64}, 1943 (1990).

\bibitem{xie} Y. H. Xie, G. H. Gilmer, C. Roland, P. J. Silverman,
S. K. Buratto, J. Y. Cheng, E. A. Fitzgerald, A. R. Kortan, 
S. Schuppler, M. A. Marcus, and P. H. Citrin, Phys. Rev. Lett. 
{\bf 73}, 3006 (1994).

\bibitem{walther} T. Walther, C. J. Humphreys, and A. G. Cullis, Appl.
Phys. Lett. {\bf 71}, 809 (1997).

\bibitem{perovic} D. D. Perovi\'{c}, B. Bahierathan, H. Lafontaine,
D. C. Houghton, and D. W. McComb, Physica A {\bf 239}, 11 (1997).

\bibitem{lagally} P. Sutter and M. G. Lagally, Phys. Rev. Lett. {\bf 84}, 
4637 (2000).

\bibitem{tromp} R. M. Tromp, F. M. Ross, and M. C. Reuter, Phys. Rev. 
Lett. {\bf 84}, 4641 (2000).

\bibitem{snyder} C. W. Snyder, B. G. Orr, D. Kessler, and L. M. Sander,
Phys. Rev. Lett. {\bf 66}, 3032 (1991).

\bibitem{gendry} M. Gendry, G. Grenet, Y. Robach, P. Krapf, L. Porte,
and G. Hollinger, Phys. Rev. B {\bf 56}, 9271 (1995).

\bibitem{guyer00} J. E. Guyer, S. A. Barnett, and P. W. Voorhees, J. 
Crystal Growth {\bf 217}, 1 (2000).

\bibitem{okada} T. Okada, G. C. Weatherly, and D. W. McComb, J. Appl. Phys.
{\bf 81}, 2185 (1997).

\bibitem{peiro} F. Peir\'{o}, A. Cornet, J. R. Morante, A. Georgakilas, 
C. Wood, and A. Christou, Appl. Phys. Lett. {\bf 66}, 2391 (1995).

\bibitem{gonzalez} D. Gonz\'{a}lez, G. Arag\'{o}n, D. Ara\'{u}jo,
and R. Garcia, Appl. Phys. Lett. {\bf 76}, 3236 (2000).

\bibitem{cahn}  J. W. Cahn, Acta. Metall. {\bf 9}, 795 (1961); Trans. 
Metall. Soc. AIME {\bf 242}, 166 (1968).

\bibitem{larche} F. C. Larch\'{e} and J. W. Cahn, Acta Metall. {\bf 33},
331 (1985).

\bibitem{guyer96} J. E. Guyer and P. W. Voorhees, Phys. Rev. B {\bf 54},
11710 (1996); Mat. Res. Soc. Symp. Proc. {\bf 399}, 351 (1996); 
J. Cryst. Growth {\bf 187}, 150 (1998).

\bibitem{spencer00} B. J. Spencer, P. W. Voorhees, and J. Tersoff, Phys.
Rev. Lett. {\bf 84}, 2449 (2000). 

\bibitem{spencer01} B. J. Spencer, P. W. Voorhees, and J. Tersoff, 
Appl. Phys. Lett. {\bf 76}, 3022 (2000); Phys. Rev. B {\bf 64}, 
235318 (2001).

\bibitem{tersoff}  P. Venezuela and J. Tersoff, Phys. Rev. B {\bf 58}, 
10871 (1998).

\bibitem{leonard98} F. L\'{e}onard and R. C. Desai, Phys. Rev. B {\bf 57},
4805 (1998).

\bibitem{leonard99} F. L\'{e}onard and R. C. Desai, Appl. Phys. Lett. 
{\bf 74}, 40 (1999).

\bibitem{huang} Z. F. Huang and R. C. Desai, submitted to Phys. Rev. B.

\bibitem{glas} F. Glas, Phys. Rev. B {\bf 55}, 11277 (1997).

\bibitem{kumar} P. Kumar and K. Maki, Phys. Rev. B {\bf 13}, 2011 (1976).

\bibitem{mullins} W. W. Mullins, J. Appl. Phys. {\bf 28}, 333 (1957).

\bibitem{chen} K. M. Chen, D. E. Jesson, S. J. Pennycook, T. Thundat,
and R. J. Warmack, J. Vac. Sci. Technol. B {\bf 14}, 2199 (1996).

\bibitem{ozkan} C. S. Ozkan, W. D. Nix, and H. Gao, Mat. Res. Soc.
Symp. Proc. {\bf 440}, 323 (1997).

\bibitem{spencer92} B. J. Spencer, P. W. Voorhees, S. H. Davis, and
G. B. McFadden, Acta Metall. Mater. {\bf 40}, 1599 (1992).

\end{thebibliography}
\end{document}